\documentclass[10pt,ngerman,oneside,onecolumn,a4paper]{article}
\usepackage{graphicx}
\usepackage[utf8]{inputenc}
\usepackage{mathptmx}
\usepackage{tabularx}
\usepackage{ragged2e}
\usepackage[singlelinecheck=false]{caption}
\usepackage[T1]{fontenc}
\usepackage[numbers,sort&compress]{natbib}
\usepackage{amsmath}
\usepackage{siunitx}
\usepackage{enumitem}
\setitemize{noitemsep,topsep=2pt,parsep=0pt,partopsep=0pt}
\usepackage{url}
\RequirePackage[blocks]{authblk}
\usepackage{babel}
\makeatletter
\let\ps@plain\ps@empty
\def\@xivpt{14bp}

\def\@sect#1#2#3#4#5#6[#7]#8{%
  \ifnum #2>\c@secnumdepth
    \let\@svsec\@empty
  \else
    \refstepcounter{#1}%
    \protected@edef\@svsec{%
      \ifnum #2<4
        \hb@xt@10mm{\csname the#1\endcsname}\relax
      \else
        \hb@xt@12mm{\csname the#1\endcsname}\relax
      \fi}%
  \fi
  \@tempskipa #5\relax
  \ifdim \@tempskipa>\z@
    \begingroup
      #6{%
        \@hangfrom{\hskip #3\relax\@svsec}%
          \interlinepenalty \@M #8\@@par}%
    \endgroup
    \csname #1mark\endcsname{#7}%
    \addcontentsline{toc}{#1}{%
      \ifnum #2>\c@secnumdepth \else
        \protect\numberline{\csname the#1\endcsname}%
      \fi
      #7}%
  \else
    \def\@svsechd{%
      #6{\hskip #3\relax
      \@svsec #8}%
      \csname #1mark\endcsname{#7}%
      \addcontentsline{toc}{#1}{%
        \ifnum #2>\c@secnumdepth \else
          \protect\numberline{\csname the#1\endcsname}%
        \fi
        #7}}%
  \fi
  \@xsect{#5}}
\renewcommand\LARGE{\@setfontsize\LARGE{16}{20}}
\def\abstract#1{\def\@abstract{#1}}
\def\abstractEn#1{\def\@abstractEn{#1}}
\def\titleEn#1{\def\@titleEn{#1}}
\def\@maketitle{%
  \newpage
  \null
  \let \footnote \thanks
    {\LARGE\bfseries\RaggedRight \@titleEn \par}%
    {\LARGE\bfseries\RaggedRight \@title \par}%
    \vskip 1\baselineskip%
    {\normalsize
      \@author\par}%
    \vskip 2\baselineskip%
    {\section*{Kurzfassung}
      \@abstract}%
    \vskip \baselineskip%
    {\section*{Abstract}
      \@abstractEn}%
  \par
  \vskip 3\baselineskip}

\renewcommand\section{\@startsection {section}{1}{\z@}%
                                   {-3.5ex \@plus -1ex \@minus -.2ex}%
                                   {\baselineskip}%
                                   {\normalfont\Large\bfseries\RaggedRight}}
\renewcommand\subsection{\@startsection{subsection}{2}{\z@}%
                                     {\baselineskip}%
                                     {1ex}%
                                     {\normalfont\large\bfseries\RaggedRight}}
\renewcommand\subsubsection{\@startsection{subsubsection}{3}{\z@}%
                                     {1\baselineskip}%
                                     {3bp}%
                                     {\normalfont\normalsize\bfseries\RaggedRight}}
\renewcommand\paragraph{\@startsection{paragraph}{4}{\z@}%
                                    {1\baselineskip\@plus1ex \@minus.2ex}%
                                    {3bp}%
                                    {\normalfont\normalsize\RaggedRight}}
\renewcommand\subparagraph{\@startsection{subparagraph}{5}{\parindent}%
                                       {3.25ex \@plus1ex \@minus .2ex}%
                                       {-1em}%
                                      {\normalfont\normalsize\bfseries\RaggedRight}}
\parindent\p@
\makeatother
\DeclareCaptionLabelSeparator{enskip}{\enskip}
\title{Drahtloskommunikation f\"ur modulare Produktionsst\"atten}
\titleEn{Wireless Communication for Modular Production Facilities}
\author{Christian Schellenberger, Marc Zimmermann, Hans D. Schotten \thanks{The authors acknowledge the financial support by the Federal Ministry for Economic Affairs and Energy of Germany in the project IC4F (grant no. 01MA17008).}}
\affil{Technische Universit\"at Kaiserslautern, Kaiserslautern, Deutschland\\ \{schellenberger, zimmermann, schotten\}@eit.uni-kl.de}

\abstract{Die wachsende Nachfrage nach individualisierten Produkten führt dazu, dass die Produktion von Gütern flexibler werden muss. Diese Veröffentlichung zeigt die Anwendungsfälle auf, aus denen sich eine modulare Produktionsstätte ableitet und zeigt die Anforderungen an die Drahtloskommunikation auf. Bereits verfügbare Technologien wie Bluetooth (classic und LE), ZigBee, IEEE 802.11 (n und ah) und 4G Mobilfunktechniken (3GPP release 8, eMTC und NB-IoT) werden ausgewertet. Weiterhin werden mögliche zukünftige Technologien wie 5G NR und private Mobilfunknetze diskutiert.}

\abstractEn{The growing demand for individualized products leads to need for a more flexible production of goods. This paper introduces the use cases that make up modular production facilities and states their requirements. Already available technologies like Bluetooth (classic and LE), ZigBee, IEEE 802.11 (n and ah) and 4G cellular technologies (3GPP release 8, eMTC and NB-IoT) are evaluated. Furthermore, possible future technologies like 5G NR and private cellular networks are discussed.}

\begin{document}
\maketitle
\section{Introduction}
The demand for individualized products is ever growing. Therefore, future production facilities have to be able to meet these demands. Linear production lines cannot meet this requirement as more time consuming processing steps for one item will jam the rest of the production line. This leads to the introduction of modular production facilities, where the unfinished products flow freely through the factory \cite{Lau16}. {\let\thefootnote\relax\footnote{{This is a preprint, the full paper has been accepted by 23th VDE/ITG Conference on Mobile Communication (23.  VDE/ITG Fachtagung Mobilkommunikation), Osnabrück, May 2018}}}
\par The goal of this paper is to present the challenges for the communication infrastructure in modular production facilities with a focus on wireless communication. The future factory will consist of many small processing entities. These can function independent of each other, like CNC machining centers, and work in parallel or serial to handle different products \cite{ZU10}. 
In order to utilize the modular production concept, the products have to be able to move freely between these different processing entities. Traditionally, the products are moving on conveyor belts, but for modular processing this is not feasible. To solve this problem automated guided vehicles (AGVs) have been introduced into factories to transport workpieces from one processing location to the next \cite{SD13}. There are two ways to coordinate the product flow through the factory and the processing of the workpieces. It can be done centralized or decentralized \cite{LD06}. In a centralized environment there is one entity that knows the status of every processing entity, AGV and workpiece, and decides where every workpiece has to be transported next and what has to be done with it there. The central entity has to communicate wirelessly with mobile nodes while permanently installed facilities can have a wired connection. In a decentralized environment on the other hand the workpiece itself or its workpiece carrier has to know which production step can be executed next and asks the processing entities and AGVs if they are available and initiates the transportation and processing accordingly. Not only the mobile nodes like AGVs and the workpiece carriers, but also non-moving participating entities, like warehouses and machining centers need to be equipped with wireless communication capabilities \cite{ZU10}. In this paper we will focus on the decentralized approach where a large number of devices have to communicate wirelessly. This leads to a more demanding communication infrastructure than in traditional production facilities.
\par The rest of the paper is organized as follows. Section \ref{sec:requirements} introduces the use cases that comprise modular production and specifies the requirements for the wireless communication. The currently available wireless communication technologies that could meet the requirements are discussed in section \ref{sec:curr_tech} and future technologies are reviewed in section \ref{sec:future_tech}. Section \ref{sec:conclusion} concludes the paper.
\section{Requirements for Wireless Communication}
\label{sec:requirements}
The concept of modular production facilities brings new use cases and, therefore, requirements into industrial production, especially for wireless communication. At the moment wireless communication is used rarely due to missing guarantees for packet delivery within a certain time. Another problem is that factory owners are very skeptical in regards to privacy protection when transmitting confidential data through a medium everyone can access. But wireless communication has advantages like saving cabling costs and the flexibility to integrate new devices wherever you want. In the following section, topical use cases, from several industry and communication associations \cite{3GPPCAV} \cite{VDE17Funk} \cite{5GPPFotF} and their associated requirements are combined and explained.
\subsection{Flexible, Modular Production Systems}
Modular production facilities have the ability to improve the flexibility, versatility and efficiency of industrial production \cite{3GPPCAV}. Small production lines can be transported and re-arranged all over the world to reduce either distance to resources or customers. Additionally, they can be combined to small linear production lines to improve the capacity in connected production steps. In order to transport workpieces from station to station, small workpiece carriers are needed to pick up the unfinished products and deliver them to the following module.
\par To exchange important information with each other about the status of the process and the time needed until the product is finished, modules and mobile carriers have to be able to communicate directly with each other or indirectly via a central infrastructure \cite{Win13}. In case of traditional static machines, the central points are connected via wire but an increasing number of devices in Industry 4.0 scenarios will communicate wirelessly due to their moving nature. Therefore, high requirements on cycle time, \SI{4}{\ms} for cyclic and \SI{10}{\ms} for acyclic data, and synchronicity (\SI{1}{\us}) are made \cite{3GPPCAV}. It is important to fulfill those demands in combination with a service availability of 99,9999\% and a data rate of up to 10\,Mb/s \cite{3GPPService} to enable coordination between cooperating entities. For the exchange of status information that is not time critical, less strict values can be accepted. In case of machine interaction, safety conditions have to be met.
\subsection{Mobile Robots}
In order to deliver resources just in time to production modules and to pick up finished goods, mobile robots are used. Small mobile carriers are moving unfinished products from station to station. Bigger AGVs can transport up to multiple tons of steel or liquids or even whole 40foot intermodal containers from point A to B \cite{3GPPCAV}. The automated transport increases the flexibility in the production process where no worker has to be called to deliver resources, saving a lot of time and storage space. Additional fork lifts and cranes complete the vehicle fleet for industrial and intra-logistics environments. The robots do not only transport goods, but also take part in monitoring their environment with, e.g. cameras to help with inventory reports and maintenance \cite{3GPPCAV}.
\par Similar to modular production systems the requirements for communication technologies are very strict. The demands for cycle time is dependent on the use case \cite{3GPPCAV}:
\begin{itemize}
	\item \SI{1}{\ms} for cooperative robotic control
	\item 1-\SI{10}{\ms} for machine control
	\item 10-\SI{50}{\ms} for cooperative driving
	\item 10-\SI{100}{\ms} for video operated remote control
	\item 40-\SI{500}{\ms} for traffic management and support systems
\end{itemize}
Additionally, it has to be taken into consideration that robotic control for cooperative driving is based on device-to-device communication, but remote control and traffic management is communication between the vehicles and the infrastructure. Therefore, a freely configurable topology is needed. Similar to the modular production use case the required service availability is 99,9999\% \cite{3GPPCAV}\cite{VDE17Funk}. The minimum data rate is 10\,Mb/s \cite{3GPPService} for video streaming and real-time data. A movement speed of 36\,km/h \cite{VDE17Funk} to 50\,km/h \cite{5GPPFotF} has to be supported. Another requirement that often accompanies the movement speed is the handover mechanism. In infrastructure communication the switch between access points has to be seamless without any loss of data or delay, especially for time-critical messages.
\subsection{Massive Wireless Networks}
In addition to the mentioned production modules and mobile robots, huge amounts of wireless sensor device, e.g. workpiece carriers, will be added into industrial environments \cite{3GPPCAV}. These sensors are able to monitor various types of parameters like volume, toxicity, pressure, acceleration or temperature. The measured value can be uploaded to either a central instance like a cloud service or processed locally. 
\par Dependent on the usage of the data, different requirements have to be addressed. Sensor data can not only be used for monitoring the production throughput (event-based), but also for maintenance purposes (interval-based) and even continuous safety critical monitoring. Condition monitoring for safety has the strictest requirements with an expected end-to-end latency of 5-\SI{10}{\ms} compared to up to \SI{1}{\s} for event-/interval-based monitoring \cite{3GPPCAV}\cite{VDE17Funk} . Service availability has to be better then 99,9999999\% for safety critical usage, but only 99,999\% for event-driven applications \cite{VDE17Funk}. Due to the huge amount of overall connected wireless devices within a production facility, a connection density of over $10^{6}$ per $km^{2}$ \cite{3GPPCAV} has to be supported. Critical messages have to be prioritized and need a guaranteed latency and reliability. The summarized data rate can go up to 100\,Mb/s \cite{3GPPCAV} that has to be routed through the network. A challenge for wirelessly connected devices in the industrial environment is the limited energy. Considering sensors that are attached to the ceiling or inside of tanks and machines, the battery lifetime should be long enough to minimize the labor cost involved in charging batteries or replacing devices. The goal should be to achieve the longest possible lifetime.
\section{Current Technologies}
\label{sec:curr_tech}
Wireless communication was traditionally only used to enable access to the company's intranet to retrieve information, e.g. for repairs, and for sensors that could not be connected by wire, like on rotating equipment. To support modular production, the new infrastructure has to satisfy the requirements stated in section \ref{sec:requirements}.
\par Some wireless protocols that have been studied extensively for the use in wireless sensor networks (WSN) are Wi-Fi \cite{802.11n}, Bluetooth \cite{802.15.1} and ZigBee \cite{802.15.4}. \cite{LS07} compares these technologies on different key aspects like range, number of nodes per access point and energy consumption. For this paper we will also study Bluetooth low energy with mesh network capabilities, IEEE 802.11ah and 802.11n, and ZigBee, as these technologies seem to be able to meet a lot of the requirements stated in section \ref{sec:requirements}. Furthermore, we will have a look at the cellular networking technologies LTE (3GPP Release 8 \cite{3GPPRel8}), enhanced machine type communication (eMTC) and narrowband IoT (NB-IoT) which were introduced with the 3GPP Release 13 \cite{3GPPRel13}.
\par For AGVs there is not yet much work done in evaluating of possible communication technologies. This is due to a prevalence of AGVs following fixed guide paths which don't need to communicate. Most works focus on these topics \cite{LD06}\cite{VI06}. In contrast, free flowing AGVs, required for modular production, need to wirelessly communicate to other entities. We will evaluate the aforementioned technologies for this use case.
\par The availability of the communication components is very important for industrial applications. For consumers communication downtime is often just unpleasant, but in an industrial environment it can lead to production delays or shut down the production temporarily. The availability of the discussed communication technologies is not well studied as it depends strongly on the implementation of the chip manufacturer and the manufacturer of the device itself. This aspect of the requirements is therefore not discussed any further in this work.
\par The key aspects of all aforementioned technologies are shown in table \ref{tab:wicomtech}. 
\begin{table*}[h]
	\caption{Comparing different wireless communication technologies \cite{LS07,3GPPRel8,3GPPRel13,AB14,ZO14,BL17,NokiaIoT,CH17,GOP12}}
	\label{tab:wicomtech}
	\begin{tabularx}{\hsize}{ p{2cm}|X|X|X|X|X|X|X|X }
		& \multicolumn{2}{c|}{Bluetooth} & \centering{ZigBee} & \multicolumn{2}{c|}{Wi-Fi} & \multicolumn{3}{c}{LTE} \\\hline
		Standar-dization & IEEE 802.15.1 Classic & IEEE 802.15.1 LE & IEEE 802.15.4 & IEEE 802.11 a/b/g/n & IEEE 802.11 ah & 3GPP Rel. 8 & 3GPP Rel. 13 eMTC & 3GPP Rel. 13 NB-IoT\\\hline
		frequency band (europe) & \SI{2.4}{\GHz} & \SI{2.4}{\GHz} & \SI{868}{\MHz}/ \SI{2.4}{\GHz} & \SI{2.4}{\GHz}/ \SI{5}{\GHz} & \SI{868}{\MHz} & \multicolumn{3}{c}{\SI{800}{\MHz}/ \SI{1.8}{\GHz}/ \SI{2.6}{\GHz}}\\\hline
		Nominal range & \SI{10}{\m} & \SI{10}{\m} & \SI{100}{\m} & \SI{100}{\m} & \SI{1}{\km} & \SI{5}{\km} & \SI{5}{\km} & \SI{8}{\km}\\\hline
		max. nodes per cell & 8 & 32767 & 65536 & 2007 & 8191 & 400 & >50000 & >50000 \\\hline
		max. data rate & 1\,Mb/s & 2\,Mb/s & 250\,Kb/s/ 20\,Kb/s & 150\,Mb/s & 7.8\,Mb/s & 100\,Mb/s & 1\,Mb/s & 250\,Kb/s \\\hline
		latency & <\SI{100}{\ms} & <\SI{3}{\ms} & <\SI{5}{\ms} & \SI{25}{\ms} &  & \SI{20}{\ms} & 10-\SI{15}{\ms} & 1.6-\SI{10}{\s} \\\hline
		Topology & P2P, star & P2P, star, mesh & star, tree, mesh & star, tree mesh, P2P & star, tree & star & star & star
	\end{tabularx}
\end{table*}
\subsection{Bluetooth}
Bluetooth classic was intended for short-range, low power devices and was mainly used for audio transmission, like hands free headsets for mobile phones and to connect to wireless speakers. The limitation of 8 active devices per (pico) cell limits the usability in production facilities drastically. With the introduction of Bluetooth low energy (LE) and the mesh structure, Bluetooth now supports up to 32767 devices per cell. With mesh capabilities there is no need for a large number of access points. The downside is that nodes close to the access points can have their batteries drained fast by relaying messages of the multi-hop communication. The upside of the wide range of supported network topologies is the flexibility of organizing the nodes as needed without the need for further access points. Due to its application in consumer electronics the reputation of Bluetooth is not very good and therefore the acceptance in the industry not very high. Furthermore, the support for low energy and mesh are not very widespread yet and the range is not very high. Another drawback of Bluetooth is the use of the \SI{2.4}{\GHz} ISM band. Due to the proliferation of wireless communication the limited license free spectrum can get crowded and regular traffic in office Wi-Fis could interfere with the factory Bluetooth network.
\subsection{ZigBee}
ZigBee is a protocol intended for embedded devices with low power consumption and low cost. In the \SI{2.4}{\GHz} frequency band it has data rates of up to 250\,Kb/s, while in the \SI{868}{\MHz} frequency band it only has 20\,Kb/s. The topologies supported by ZigBee are star, tree and mesh and it is therefore quite flexible. With a higher nominal range than Bluetooth the the number of access points can be lower. The higher nominal range furthermore requires less hops for the same distance which reduces the burden on the batteries of stations that would otherwise be relaying messages. The support of 65536 devices per cell is the highest of the current technologies that we examined. As mentioned in the previous subsection the frequency band of \SI{2.4}{\GHz} is quite crowded and could therefore lead to interferences, while the less used \SI{868}{\MHz} frequency band has a very low data rate compared to the other technologies.
\subsection{Wi-Fi}
Consumer oriented IEEE 802.11n \cite{802.11n} is not intended for machine to machine communication (M2M), but for inter computer networking. These networks have very high data rates of 150\,Mb/s (single antenna), but a very high overall energy consumption. Another limitation is the number of nodes per cell. 2007 nodes can be too low if intelligent workpiece carriers are stored very tightly. The usage of multiple access points can increase the number of supported nodes, but it leads to interferences due to a limited number of communication channels. Therefore, an amendment to the standard was made in the form of 802.11ah \cite{802.11ah}. The energy consumption was reduced drastically. This is done by shortening the transmission time with a shortened header and null data packet frames to shorten signaling frames like ACKs and CTS. Through the usage of the \SI{868}{\MHz} frequency bands the nominal range extends to \SI{1}{\km} and the number of access points can be reduced, if the node density is not very high. Furthermore, 802.11ah is operated in a frequency band that is not used much for consumer applications and therefore less interference from other regular traffic is expected. This is the same frequency band as can be used with ZigBee, but 802.11ah reaches far higher data rates. Adame et al. expect the use of 802.11ah as a legacy solution for the deployment of WSNs in most markets. While the Bluetooth and ZigBee incorporate methods to guarantee latencies the standard medium access technology of 802.11n does not. The traffic indication map (TIM) of 802.11ah together with the restriced access window can be defined in a way that every station has its own time slot resulting in a guaranteed latency.
\subsection{LTE}
The 4th generation (4G) cellular communication standard, named longterm evolution (LTE), was introduced with 3GPP release 8 in 2008 \cite{3GPPRel8}. It significantly improved the data throughput of mobile networks to cope with the growing demand for high data rates on mobile phones. This version of LTE does only support up to 400 nodes per cell, which is not enough for high density wireless sensor networks required for a modular production facility. In release 13 two M2M communication specifications were introduced to natively support WSN applications. One is narrowband IoT (NB-IoT) and the other is enhanced Machine Type Communication (eMTC). They both support over 50000 nodes per cell and have a range of \SI{5}{\km} (eMTC) and \SI{8}{\km} (NB-IoT). However, there are major differences. NB-IoT only supports stationary nodes while eMTC supports the handover between cells. While NB-IoT is only half duplex, eMTC also has a full duplex option. eMTC has a latency of 10-\SI{15}{\ms} and supports voice transmission, while NB-IoT has a latency of 1.6-\SI{10}{\s} and does not support voice transmission. On the other hand, NB-IoT chips are expected to be cheaper and more energy efficient. One upside of NB-IoT is its capability to re-farm the GSM spectrum. A problem concerning the security of the transmitted data is that cellular networks are operated by telecommunication providers and therefore all traffic has to be routed through their networks. The transmission through public networks furthermore increases the end to end latency of the communication. This is major drawback as industry companies want to control where their data is moving. One possibility to solve this problem is the introduction of private networks which is discussed in section \ref{Priv_Netw}.
\subsection{Evaluation of the current technologies}
The different use cases that are combined for the modular factory differ in their requirements. Not one of the compared technologies can meet all requirements. Several technologies can meet the requirements for massive wireless networks when the nodes are not used for safety applications. Only Bluetooth LE and ZigBee can meet the 5-\SI{10}{\ms} latency required. The \SI{1}{\s} requirement for monitoring can be met by 802.11ah. Bluetooth Classic and LTE are not usable for massive sensor networks as the number of supported nodes per cell is to low. For AGVs a data rate of at least 10\,Mb/s is required which can only be achieved by 802.11n or LTE networks. With the routing of data through public networks, LTE is not the preferred choice for factory operators, but can be the better solution if a company has large outdoor areas, which are costly to equip with the high number of required IEEE 802.11n access points. Another advantage of IEEE 802.11n compared to LTE is the capability of using ad-hoc or point-to-point (P2P) networks. AGVs could use ad-hoc networks with surrounding AGVs in order to cut down latency and improve data transmission rates, which is useful for cooperative tasks. This leads to need for heterogeneous wireless technologies in order to implement modular production facilities with currently available technologies.
\section{Future Technologies}
\label{sec:future_tech}
It is expected that 5G new radio in conjunction with private networks can meet or even exceed the requirements stated in section \ref{sec:requirements} with one technology. 
These two technologies and concepts will be discussed in this section.
\subsection{5G New Radio}
As the new mobile communication generation, called 5G, is promising to be the key enabler for the next level of wireless communication \cite{5GPPFotF}\cite{BMVI5G}\cite{ZVEI5G}, it is important to evaluate it for the usage in the context of Industry 4.0. Due to its heterogeneity it is supposed to support nearly all uses cases for industrial applications \cite{3GPPCAV}. It promises to cover extreme mobile broadband (xMBB) requirements with data rates up to 10\,Gb/s, massive machine-type communication (mMTC) with 1 million devices per $km^{2}$ and even ultra-reliable (>99,9999\%), low latency (<1ms) communication (uMTC and URLLC) for time-critical closed-loop applications \cite{VDE17Funk}\cite{NOK5G}. With its scalability and customizability, users should be able to cover their needs with one technology that acts as a platform for interconnecting machines, robots, processes, AGVs, goods and workers \cite{5GPPFotF}. The function as a platform will blur the boundaries between wide-, local, and personal-area networks. One requirement for 5G is the  seamless connectivity to coexisting access technologies. 5G also has to incorporate efficient energy saving strategies to enable battery lifetimes up to 10 years with low-energy communication.
\par To achieve all those different requirements that partially exclude each other, like high data rates and low latency, several technologies have been introduced. 5G is using re-configurable radios to switch between frequency bands, physical layer parameters, like symbol duration, and diversity in time, space and frequency \cite{NOK5G}. Its flexibility is also reached by usage of network slicing where multiple logical networks with different capabilities can be realized on one physical net \cite{VDE17Funk}. This will be enabled by the usage of software defined networking (SDN) and network function virtualization (NFV). SDN uses a controller that decides on the specific demand for an incoming communication request which of the available logical networks is suitable. With NFV network services and whole network classes can be virtualized to create new communication services. Those technologies enable operators to use their physical infrastructure to create new virtual instances of an entire network. With network slicing a manageable Quality of Service (QoS) and Quality of Experience (QoE) \cite{NOK5G} can be achieved. The difference between QoS and QoE is that with QoE, errors may occur, but the application or the user doesn’t notice them, e.g., frame drop on video streaming. Additionally, communication groups with restricted access can be created to increase security and ensure privacy.
\subsection{Private Cellular Networks}
\label{Priv_Netw}
The advantage of using non cellular communication is that operators can equip their factories with their own infrastructure and directly connect it to their internal networks. At the moment only telecommunication providers can operate cellular networks as the spectrum that the technologies are used in have to be licensed nationwide by regulatory bodies like the BNetzA in Germany. For the future, there are plans for factory owners to install their own cellular networks through new regulations enabling private 4G/5G cellular networks.
\par A lot of the previously listed requirements may not be achievable with the use of the public access networks used for current mobile communication. The main problem is that data is sent to the provider’s base station and routed through a public network and delivered to the destination that might only be some meters away from the source when focusing on industrial communication. Not only does the unnecessary long way and the delay from routing procedures hinder low latencies, but also potentially violates privacy policies. Therefore, 5G NR and the current 4G/LTE-A, should support private networks. Those private networks consist of a private base station that is owned and managed by the companies themselves which leads to individualized configurations for latency requirements and QoS constraints.
\par To realize private cellular networks national and international regulation bodies like the BNetzA in Germany need to decide whether to assign frequency bands for the local use or not. Those special frequencies might be assigned with the licensed shared access (LSA) approach \cite{ECC205}\cite{Intel15} where secondary services have a location-specific and temporary access. In Germany the frequencies around \SI{2}{\GHz} and \SI{3.6}{\GHz} have been identified to be used for this approach \cite{BNetzA17}. They can be used for either 5G \cite{Mas17} or for any other communication technology \cite{ETSI13}. In case, that they will be available after the next frequency allocation process, most likely at the beginning of 2019, private groups can buy access to those frequencies restricted in time, location and bandwidth to build their own local high-quality networks. They could use technologies now operated in licensed frequencies, like LTE, without involvement of today’s operators or implement technologies used in license-free frequencies, like Wi-Fi or Bluetooth, in less crowded licensed spectrum \cite{BNetzAAllg}.
\par Right now it is not sure, if this approach can be realized, because today’s operators fear to not be able to meet the political demand that Germany takes the leadership in the introduction of 5G NR if too much of the available bandwidth is reserved for local allocations \cite{BNetzAStellung}. They rather want to lease locally restricted frequencies to customers or offer them the private network as a service for an annual fee. Nevertheless, private networks are a key factor to meet most of the requirements with 5G. Even the standardization initiative 3GPP demands that 5G user equipment (UE) has to support public and private networks simultaneously and switch seamlessly between both. With this capability use cases are covered where UEs enter or leave areas where poor public coverage is compensated by good private coverage.
\section{Conclusion}
\label{sec:conclusion}
At the moment there is no one-size-fits-all wireless communication technology to satisfy all requirements for modular production facilities. There are some current technologies that can solve most of the requirements, but high bandwidth combined with very low latencies connected with the AGV use case can yet not be achieved with widely used standardized technologies. The upcoming 5G in conjunction with private cellular networks promises to meet the demanding requirements.
\bibliography{Literature}
\end{document}